Revealing degradation mechanisms in YSZ ceramics through machine learning-guided aging and multiscale characterization

Prachi Garg[1], and Baishakhi Mazumder[1]*

*1 Department of Materials Design and Innovation, State University of New York at Buffalo, USA*

*baishakh@buffalo.edu (corresponding author)*

Abstract

The long-term performance of yttria-stabilized zirconia (YSZ) based energy and biomedical devices is compromised by low-temperature degradation (LTD). This study presents a novel integration of machine learning-guided hydrothermal aging with multiscale characterization to resolve a two-stage degradation mechanism in 3 mol% YSZ. Stage 1 (0 to 30 hrs) features initial surface relief building, which transitions to partial refinement and relief distribution in stage 2 (30 to 60 hrs), alongside a rise in monoclinic phase content. The evolving microstructure increases triple-junction grain boundary density, and these junctions act as degradation hotspots, where vacancy exchange and water access accelerate the transformation. These findings highlight grain boundary chemistry, rather than grain size alone, as a key LTD driver, suggesting boundary engineering as a strategy to enhance YSZ stability for energy, biomedical, and thermal applications.

## Machine learning guided artificial aging

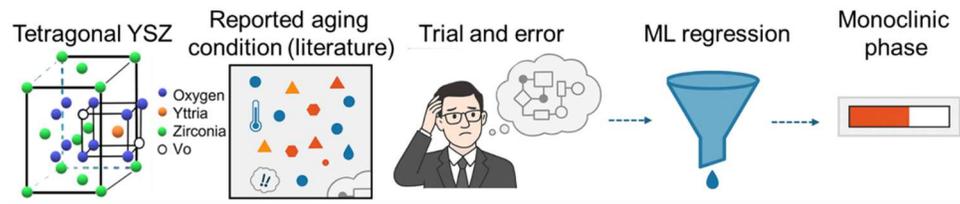

## Multiscale characterization integrated strategy

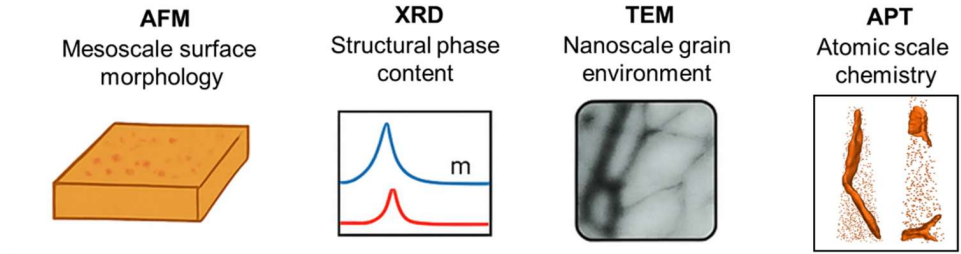

## Two stage aging mechanism

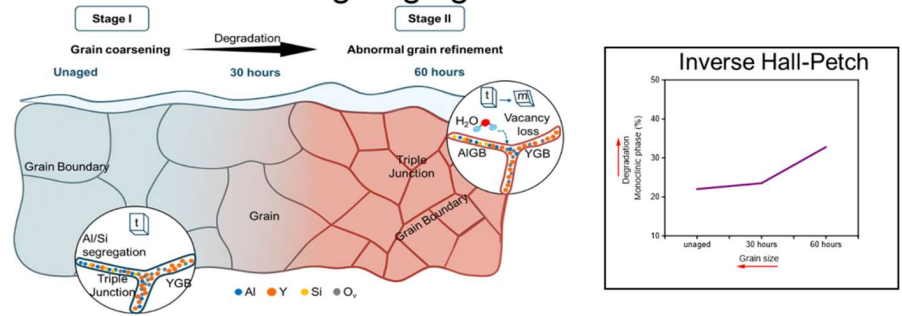

An integrated approach combining machine learning and multi-scale characterization reveals a two-stage degradation mechanism in YSZ. The top panel depicts the need for ML guidance to increase the time and cost efficiency by predicting aging parameters that influence monoclinic phase content. The middle panel shows a multi-scale experimental approach from meso to atomic-scale analysis to capture the complexity of the aging process. The bottom panel displays a two-stage degradation mechanism uncovered from our study, indicating an inverse Hall-Petch phenomenon

## 1. Introduction

Yttria stabilized zirconia (YSZ) plays a central role in advanced ceramics and is recognized for its exceptional mechanical toughness, ionic conductivity, and thermal stability. These properties fuel their widespread use in solid oxide fuel cells, thermal barrier coatings, oxygen sensors, and biomedical implants like dental restorations [1,2,3,4,5]. The key to YSZ's performance lies in its metastable tetragonal phase, where oxygen vacancies suppress the transformation to the monoclinic form. This enhances YSZ's durability via transformation toughening, a process known for over two decades, where stress-induced phase shifts create compressive barriers to crack growth [6,7,8]. Yet, this stability falters under humid, moderate-temperature conditions, triggering low-temperature degradation (LTD) [9,10]. Initiating at the surface and often nucleating at grain boundaries, LTD drives a tetragonal-to-monoclinic (t→m) transformation that leads to microcracking and diminished performance [11,12,13], posing a persistent challenge for YSZ's long-term reliability in critical applications.

While decades of research have documented LTD's key players as grain environment [14,15], dopant segregation [16,17,18], and oxygen vacancies [19,20], the key questions persist: How do grain boundaries evolve during aging? What roles do dopant segregation play? How do structural and chemical changes interact across scales? Most efforts have focused on either surface morphology or bulk phase content, often overlooking the grain boundary networks and the local chemistry that mediate aging [15,21]. This critical gap hampers the development of targeted strategies to enhance the stability of YSZ, especially for applications demanding long-term performance under harsh environments.

To bridge this gap, we conducted a systematic, multiscale study of aged 3 mol% YSZ. Leveraging over 20 years of foundational research, we developed machine learning regression models trained on literature-reported aging data to simulate and predict degradation behavior. These data-driven models guided experimental designs by optimizing aging parameters such as temperature, time, and pressure. This helped induce monoclinic phase contents ranging in tetragonal YSZ, enabling a controlled step-by-step dissection of the degradation mechanism. A major strength of this study lies in its comprehensive and innovative integrated multiscale characterization strategy to capture degradation across multiple length scales. Atomic force microscopy (AFM) was employed to probe mesoscale surface morphology, while X-ray diffraction (XRD) provided depth-dependent phase distribution. Crucially, atom probe tomography (APT) enabled atomic-scale, three-dimensional chemical mapping, enabling us to track elemental distribution and chemical evolution at grain boundaries and triple junctions, with unmatched resolution. Importantly, this study provides the first direct evidence that triple junction grain boundaries (TJGBs) act as chemically and structurally unstable hotspots, governing LTD progression; an insight not captured in previous LTD studies.

Our results uncover a two-stage degradation sequence, that results in increase of triple junction formation, amplifying sites for degradation. These sites emerge as critical zones for t→m transformation and oxygen vacancy loss, modulated by dopant and impurity segregation, which can make them potential hotspots for structural and chemical instability. Notably, despite their high energy and chemistry, triple junction grain boundaries (TJGB) are not explored in YSZ studies. By connecting structure and chemistry across length scales, this study offers new insights into LTD and a foundation for engineering next-generation YSZ-based ceramics for demanding environments.

## 2. Methodology

All the experiments were performed on square samples of 3 mol% YSZ (named as IPS e.max ZirCAD LT) obtained from Ivoclar Vivadent. This material is made by intentionally doping zirconia with yttria to stabilize the metastable tetragonal phase. No alumina was intentionally added during sample processing.

## 2.1. Machine learning model-guided aging

To study the evolution of grains over time, we mimicked the aging process in an artificial environment. Supervised machine learning models were used to predict the extent of phase degradation (i.e., monoclinic phase fraction) based on aging process conditions. To train these models, 200 data points were collected form literature including aging parameters such as temperature ($^0$C), time (hrs.), pressure (bar) and target property- monoclinic phase content (%). Data mining was done to eliminate repetitive data, resulting in ~100 data points in our input dataset (shown in supplementary file S1). Eight predictive regression models, Linear Regression (LR), Multiple Linear Regression (MLR), Principal Component Regression (PCR), Ridge Regression, Lasso Regression, Support Vector Regression (SVR), Gaussian Process Regression (GPR), and Random Forest (RF) were evaluated and trained on this dataset. To ensure the reliability of our model and its ability to handle small datasets with nonlinear trends, we applied a cross-validation method to assess its performance. Based on the root mean square error values, the GPR model was selected to optimize the aging parameters. These predictions averaged over five iterations to minimize any variance, with results comparable to a standard 5-fold cross-validation method. This approach was particularly useful in avoiding the information leakage problem that can occur in traditional k-fold cross-validation, where random splits of data may lead to biases in model performance.

## 2.2. Atomic Force Microscopy (AFM)

Grain evolution during aging was characterized by high-resolution surface imaging using Asylum Research AFM in tapping mode. AFM utilizes atomic forces to map tip-sample interactions, providing exceptional surface detail. Samples were scanned at 0, 15, 30, 45, and 60 hrs. For each aging step, $\geq 3$ independent areas were imaged to assess reproducibility, and where stage coordinates allowed, the same field of view was revisited across time points. Height images were collected at three magnifications tailored to the analysis: 20 µm (surface roughness: Rq) and 2 µm (grain-size statistics). Standard first-order plane subtraction was applied to remove sample tilt. Raw and processed images, including overlays of height maps with segmentation masks, are provided in Fig. 2. For roughness, Rq (root-mean-square roughness) was computed from the 20 µm scans using the instrument software after plane fit. For grain-size statistics, boundaries were delineated from the 2 µm height maps by detecting boundary grooves and manually tracing in ImageJ. To suppress spurious minima from noise or tip convolution, a light median filter was applied to the height map before tracing, and a maximum area cutoff corresponding to the same grain environment analysis was enforced. Grain size for each closed region was defined as the equivalent circular diameter, $D\text{eq} = 2A/\pi$, where $A$ is the segmented grain area. The resulting distributions were summarized as box plots (median and interquartile range, IQR; whiskers = 1.5×IQR; outliers plotted explicitly). Sample sizes (n) per time point are reported in the figure.

## 2.3. X-Ray Diffraction (XRD)

Phase transformation in YSZ was studied using X-ray diffraction (XRD) on unaged, 30 hrs, and 60 hrs aged samples. Standard theta-2theta scans identified the monoclinic phase (~28°) and tetragonal phase (~30°). Grazing incidence XRD (GI-XRD) with varying omega scans at 1°, 2°, and 5° was used to quantify depth-dependent phase changes across all samples. In grazing-incidence mode, incident angles just above the critical angle (e.g., 0.5-1°) result in nanometer-scale penetration depths. With higher angles (e.g., 2° or 5°), the X-ray beam penetrates more deeply, depending on the material absorption properties. These general ranges are consistent across oxide systems and standard lab based GIXRD configurations. Monoclinic phase fraction ($C_M$) was calculated using Toraya's equation, integrating intensity ratio of monoclinic (111) and (−111) peaks to the tetragonal (101) peak:

$$Vm = \frac{1.311 Xm}{1 + 0.33 Xm}$$

$$Xm = \frac{[Im(-111) + Im(111)]}{[Im(-111) + Im(111) + I(101)]}$$

where $I_m$ and $I_t$ are the integrated intensities of monoclinic and tetragonal peaks, respectively.

### 2.4. Atom Probe Tomography (APT)

Local chemical changes at grains and grain boundaries were analyzed using APT on unaged, 30 hrs, and 60 hrs aged samples, with two specimens per condition for statistical reliability. Needle-shaped tips (<100 nm radius) were prepared via FIB lift-out, mounted on silicon microtip arrays, and cleaned using low-kV milling. Data were collected with a CAMECA LEAP 4000XHR in laser-assisted mode (355 nm UV laser, 50 K, 60 pJ pulse energy, 125 kHz pulse rate, 0.5% evaporation rate). Reconstructions and analyses were performed using IVAS 3.8 software. Oxygen vacancy distribution was assessed via pseudo-coordination number (PCN) mapping, where PCN for each Y atom was calculated based on the number of oxygen neighbors among its 8 nearest neighbors, with lower oxygen counts indicating higher vacancy concentrations.

## 3. Results

### 3.1. Artificial aging simulation

To systematically dissect the aging mechanisms of 3 mol% YSZ we employed machine learning to predict hydrothermal aging conditions targeting specific monoclinic phase contents. As shown in Figure 1(a), GPR was selected due to its superior performance, achieving an R-squared value of 87% and exhibiting minimal difference between train and test RMSE, indicating a reduced risk of overfitting and ensuring a robust predictive model. The optimized conditions, predicted by the GPR model, are summarized in Figure 1(b). Samples were aged in an autoclave at 132°C and 2 bar pressures for 0, 15, 30, 45, and 60 hrs. This predictive approach enabled precise control over phase transformation, setting the stage for multiscale analysis via XRD, AFM, and APT.

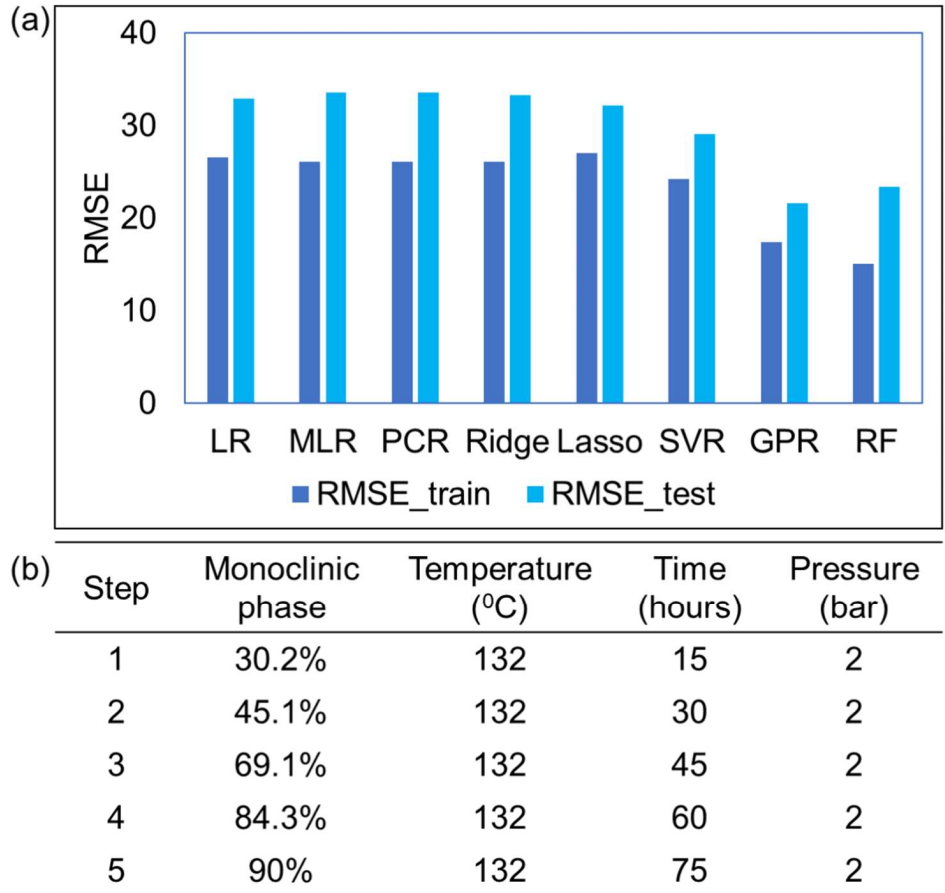

(b)

| Step | Monoclinic phase | Temperature (°C) | Time (hours) | Pressure (bar) |
|------|------------------|------------------|--------------|----------------|
| 1 | 30.2% | 132 | 15 | 2 |
| 2 | 45.1% | 132 | 30 | 2 |
| 3 | 69.1% | 132 | 45 | 2 |
| 4 | 84.3% | 132 | 60 | 2 |
| 5 | 90% | 132 | 75 | 2 |

Figure 1. (a) RMSE values for eight regression models, highlighting GPR's balanced performance with minimal overfitting, (b) Aging conditions and predictive monoclinic content

### 3.2. Surface evolution

Figure 2 depicts the grain evolution at the surface level of hydrothermally aged 3 mol% YSZ at each step. In the initial phase (0 to 30 hrs), boundary relief becomes more pronounced and larger grains appear (Fig. 2a-c). Consistently the box plot at 30 hrs exhibits the widest interquartile range and the longest upper whisker, indicating a broadened distribution with a pronounced large-grain tail; the median grain size is slightly higher than at 0 to 15 hrs. Surface roughness (Rq) shown in the table follows the same trend, rising from 115.8 nm (0 hrs) to 218.7 nm (15 hrs) and peaking at 308.3 nm (30 hrs). Beyond 30 hrs, the distribution tightens where at 45 hrs the box narrows and the upper whisker shortens, and by 60 hrs the spread remains smaller (with some residual outliers), while the median drops (Fig. 2d-e) The roughness decreases in parallel to 259.8 nm (45 hrs) and 202.6 nm (60 hrs) as shown in the table. Together, these observations identify a coarsening maximum at 30 hrs, followed by partial refinement by 45 hrs and 60 hrs.

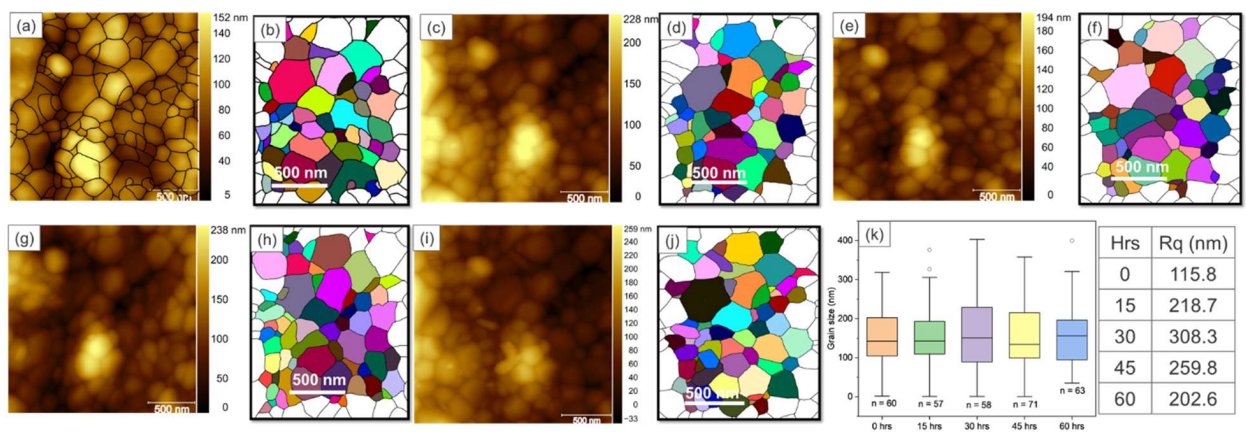

Figure 2. (a-j) Grain maps from tapping mode AFM height images at 0, 15, 30, 45, and 60 hrs (field of view 2um), illustrating grain evolution. AFM raw images of (a) 0 hrs, (c) 15 hrs aged, (e) 30 hrs aged, (g) 45 hrs aged, (i) 60 hrs aged, 3mol% Yttria stabilized zirconia are shown. The manual annotation of the grains has been shown in (a) for reference. Each annotation was used to mask the grains in image J and create grain maps for respective cycles shown in (b) 0 hrs, (d) 15 hrs, (f) 30 hrs, (h) 45 hrs, and (j) 60 hrs aged sample. (k) Grain size distribution summarized as box plots (center line is median, box is interquartile range, whiskers are 1.5 times IQR, outliers are shown). Sample counts (n) are indicated for each time point beneath the boxes. Table: surface roughness (Rq) at each step for the same area (field of view 20 um)

### 3.3. Phase transformation

To elucidate the surface microstructural evolution with crystallographic information, grazing incidence XRD (GI-XRD) at 1°, 2°, and 5° omega angles (probing surface, intermediate, and bulk regions, respectively) was performed. Figure 3 reveals the depth-dependent evolution of the tetragonal-to-monoclinic (t→m) transformation of 3 mol% YSZ during aging. In Figure 3 (a), the unaged sample, 30 hrs, and 60 hrs samples display the tetragonal peak at ~30° dominating (increasing from $1^0$ to $5^0$ across all depths, while the monoclinic peak at ~28° is only seen in 60 hrs. By 60 hrs (Figure 3 (b)), the monoclinic peak intensifies as compared to negligible in unaged and 30 hrs, indicating a measurable phase transformation progressing between 30 and 60 hrs. Figure 3 (c) depicts the monoclinic phase fraction ($C_M$) quantified via Toraya's equation. Monoclinic phase fractions at 60 hrs confirms the gradient; $C_M$ decreases as we go from the surface ($1^0$) to the bulk ($5^0$). This depth-dependent profile underscores a surface-initiated degradation process, with hydrothermal effects driving the t→m transformation from 30 hrs to 60 hrs predominantly at the material's exterior.

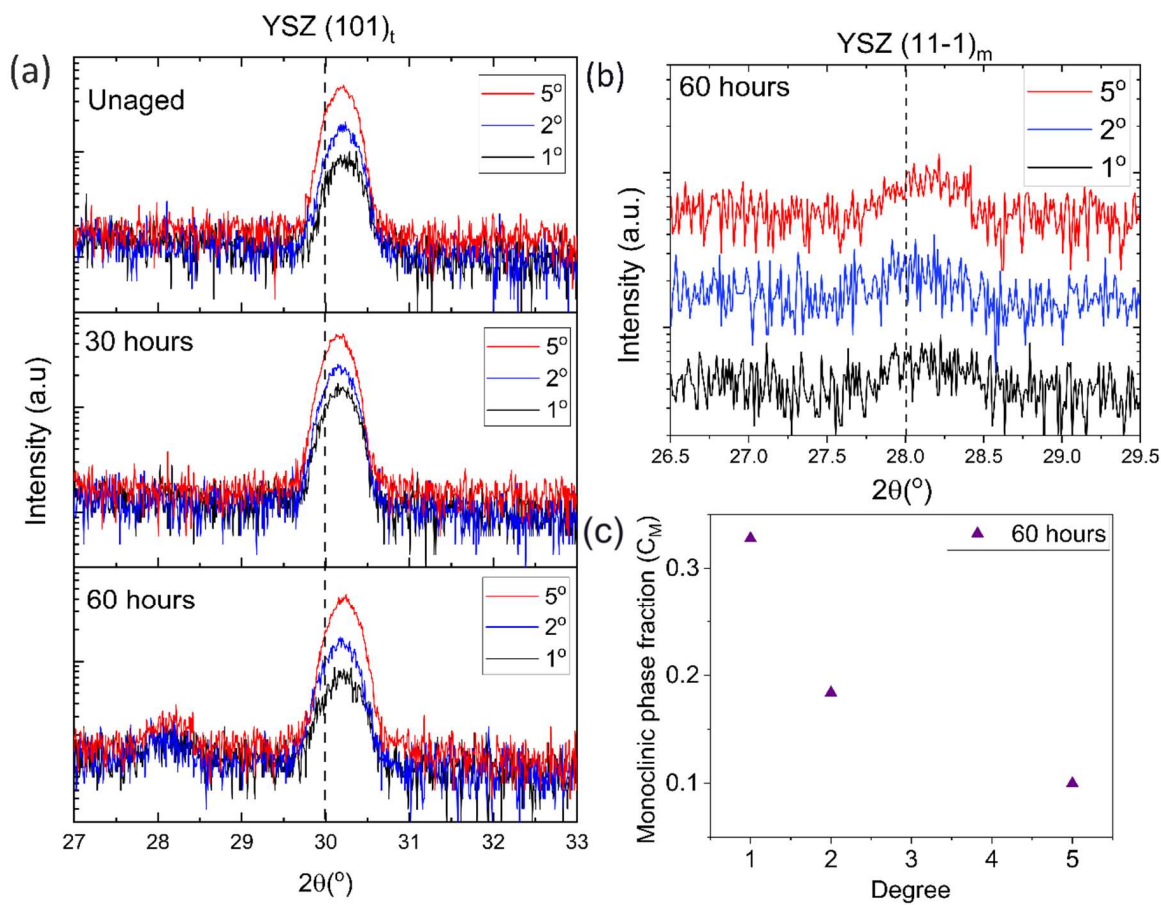

Figure 3. GI-XRD patterns of YSZ for (a) unaged, 30 hrs, and 60 hrs aged samples, showing patterns at 1° (black, surface), 2° (blue, intermediate), and 5° (red, bulk) omega angles. The monoclinic peak (m) at ~28° in 60 hrs sample (b), alongside the tetragonal peak (t) at ~30° in all samples (a) is indicated. Depth resolved monoclinic phase fraction ($C_M$) is shown in (c) by purple triangles.

### 3.4. Grain boundary chemistry and characteristics

To further enhance the understanding of spatial distribution of these cations and grain boundary network, APT was done. Atom maps (Figure 5 (a-c)) reveal yttrium (Y) segregation to grain boundaries (orange dots). The presence of zirconium (Zr), oxygen (O) was seen alongside impurities like aluminum (Al), silicon (Si), and hafnium (Hf). APT analysis of grain boundaries (GBs) in 3 mol% YSZ revealed distinct compositional changes at yttria-rich grain boundaries (YGB) and Al-rich grain boundaries (AlGB) during aging, as summarized in Figure 5 (d) and Table 1. APT measurements of GB thickness (Figure 5 (d)) provide further insight into structural evolution. YGB exhibits greater thickness (5.3nm, 4.9nm, and 3.7nm at 0, 30, and 60 hrs) than AlGB (4.3nm, 3.7nm, and 2.4nm at 0, 30, and 60 hrs across all aging conditions. This aligns well with the literature, where Koji Matsui et al. reported that $Y^{3+}$ segregation leads to increased grain boundary width compared to $Al^{3+}$ segregation. For both YGB and AlGB, the GB thickness decreases moderately from unaged to 30 hrs, with a more pronounced increase from 30 to 60 hrs.

In addition to the GB thicknesses, aging also alters boundary chemistry, as seen in Table 1. Please note all the compositions are reported in atomic percentage (at. %). For the unaged sample, AlGB shows elevated aluminum (0.37 at. %) and silicon (0.12 at. %) levels, indicating segregation of these light ions at Al-rich interfaces. After 30 hrs of aging, YGB maintains stable yttria content (2.00 at. %), but AlGB shows increased Al (0.36 at. %) and Si (0.21 at. %) segregation, alongside a rise in yttria (3.53 at. %), suggesting yttria diffusion toward Al-rich boundaries. By 60 hrs, YGB yttria content increases (3.00 at. %), while AlGB shows a slight decrease (2.63 at. %), with Al (0.39 at. %) and Si (0.24 at. %) levels peaking, reflecting continued segregation of light ions. Zirconia content remains relatively stable across all conditions (26.8-28.5 Zr at. % for YGB, 27.3-27.9 Zr at. % for AlGB), indicating that the t→m transformation primarily influences cation segregation rather than zirconia distribution. These compositional and structural trends at YGB and AlGB underscore the influence of grain boundary chemistry on surface-initiated degradation, with AlGB serving as a key site for LTD progression due to its enhanced segregation and thickening during aging.

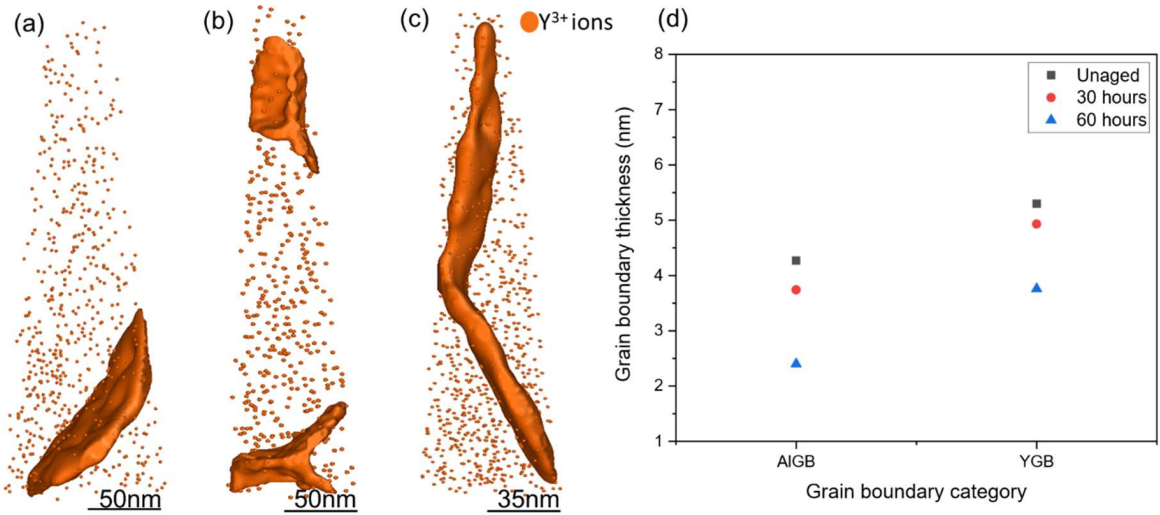

Figure 5. (a-c) Atom maps depicting Y elemental segregation across (a) unaged, (b) 30 hrs, and (c) 60 hrs aging conditions obtained via APT. (d) Grain boundary thickness of YGB and AlGB for unaged (black squares), 30 hrs (red circles), and 60 hrs (blue triangles) aged YSZ, measured via APT. (Please note only Y ions shown in (a-c) for visual clarity of grain boundary microstructure)

Table 1. Compositional analysis of YGB and AlGB in unaged, 30 hrs, and 60 hrs aged YSZ

| Composition | at. % | Unaged | | 30 hours | | 60 hours | |
|---|---|---|---|---|---|---|---|
| | | YGB | AlGB | YGB | AlGB | YGB | AlGB |
| Light ions | O | $69.9 \pm 2.7$ | $69.2 \pm 2.0$ | $69.2 \pm 0.7$ | $69.9 \pm 1.4$ | $68.1 \pm 1.1$ | $69.2 \pm 1.1$ |
| | Al | $0.04 \pm 0.0$ | $0.37 \pm 0.1$ | $0.00 \pm 0.0$ | $0.36 \pm 0.1$ | $0.07 \pm 0.0$ | $0.39 \pm 0.1$ |
| | Si | $0.04 \pm 0.0$ | $0.12 \pm 0.0$ | $0.06 \pm 0.0$ | $0.21 \pm 0.1$ | $0.02 \pm 0.0$ | $0.24 \pm 0.1$ |
| | Zr | $26.8 \pm 2.6$ | $27.9 \pm 1.4$ | $28.4 \pm 0.5$ | $25.8 \pm 2.2$ | $28.5 \pm 0.6$ | $27.3 \pm 1.1$ |
| Heavy ions | Y | $2.78 \pm 0.7$ | $2.17 \pm 0.9$ | $2.00 \pm 0.0$ | $3.53 \pm 1.1$ | $3.00 \pm 0.7$ | $2.63 \pm 0.1$ |
| | Hf | $0.21 \pm 0.1$ | $0.15 \pm 0.2$ | $0.22 \pm 0.1$ | $0.19 \pm 0.1$ | $0.25 \pm 0.1$ | $0.15 \pm 0.1$ |

Two-dimensional concentration profiles (2DCP) derived from APT (Figure 6) reveal the spatial distribution of Zr, Y, Al, Si, and O at YGB and AlGB in 3 mol% YSZ across aging conditions. In the unaged, 30-hour, and 60-hour samples (Figure 6 (a-c)), Al and Si segregation intensifies in AlGB, consistent with their faster diffusion rates compared to Y. Oxygen distribution remains relatively uniform across all conditions, suggesting that cation segregation drives structural changes at GB. This progressive segregation of Al and Si likely promotes the formation of ion-vacancy clusters [22], which may stabilize the local grain boundary structure and influence the t→m transformation during LTD.

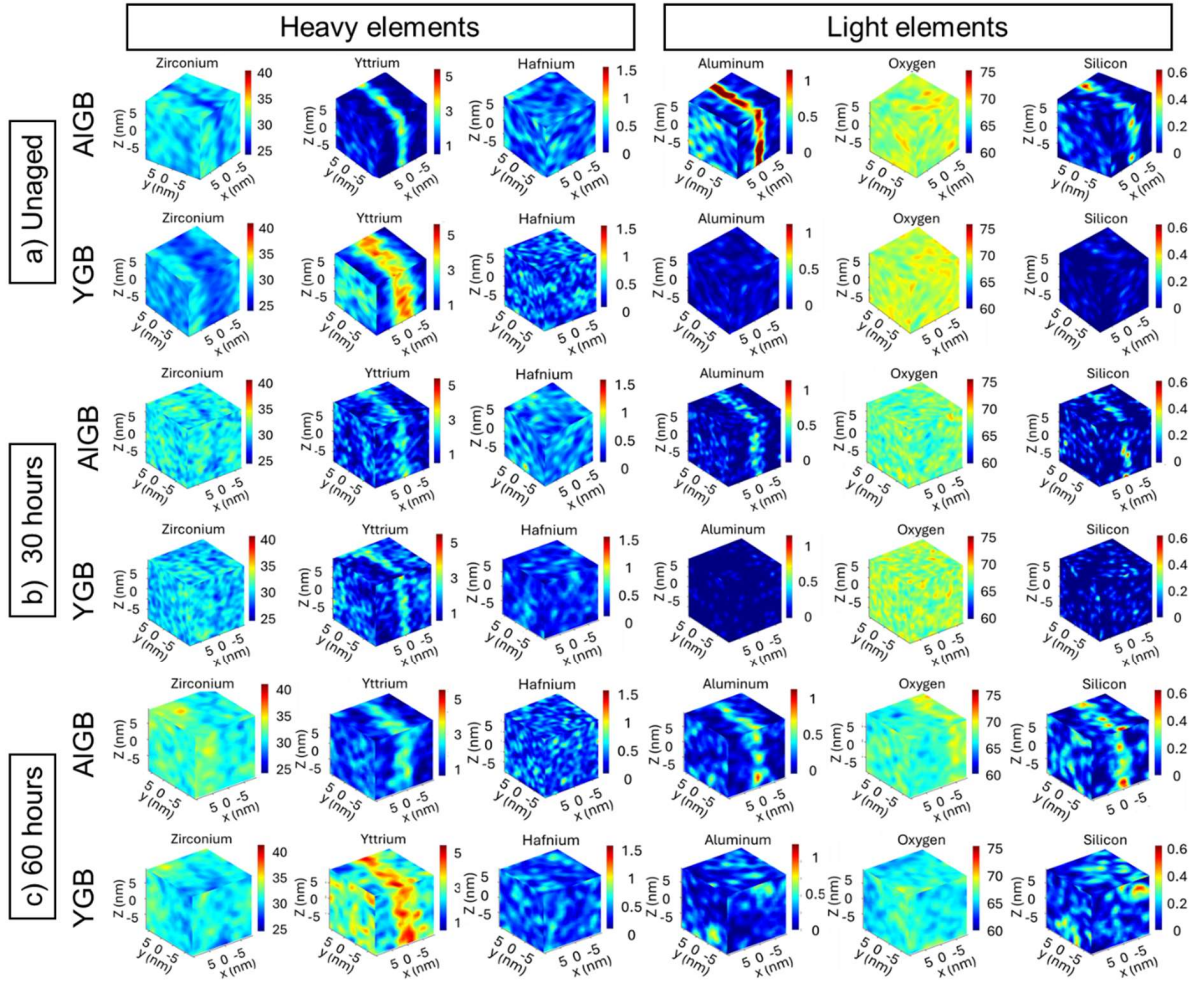

Figure 6. Two-dimensional concentration profiles of heavy elements: Zr, Y, Hf, and light elements: Al, O, Si at GBs in 3 mol% YSZ for (a) unaged, (b) 30 hrs, and (c) 60 hrs aged samples. Color scales indicate atomic concentrations (at. % %) along the x, y, and z axes (nm).

Oxygen vacancy distribution, assessed via pseudo-coordination number (PCN) (adopted from our previous work [24]), derived from APT data, reveals distinct differences between triple junction grain boundaries in 3 mol% YSZ across aging conditions. Figure 7 exhibits relative oxygen vacancy concentrations by counting the O neighborhood ($O_{NN}$) of each Y ion near the triple junction grain boundary (TJGB) region. TJGB showed Y with reduced $O_{NN}$ (Unaged: 38%, 30 hrs: 29.8%, 60 hrs: 28.7%) as compared to the bulk region (Unaged: 21%, 30 hrs: 20%, 60 hrs: 25.2%) across all aging conditions. O poor neighborhood in TJGB region, as compared to the O rich neighborhood in bulk region, indicates vacancy-rich zones in TJGB attributed to their disordered, high-energy structure. Interestingly, there is a relative vacancy reduction only in TJGB upon aging, due to water-mediated annihilation under hydrothermal conditions. Moreover, AlGB retains more vacancies (Unaged: 31%, 30 hrs: 34%, 60 hrs: 39%) as compared to YGB (Unaged: 29%, 30 hrs: 27%, 60 hrs: 29%). This reflects Al's ability to trap vacancies during aging.

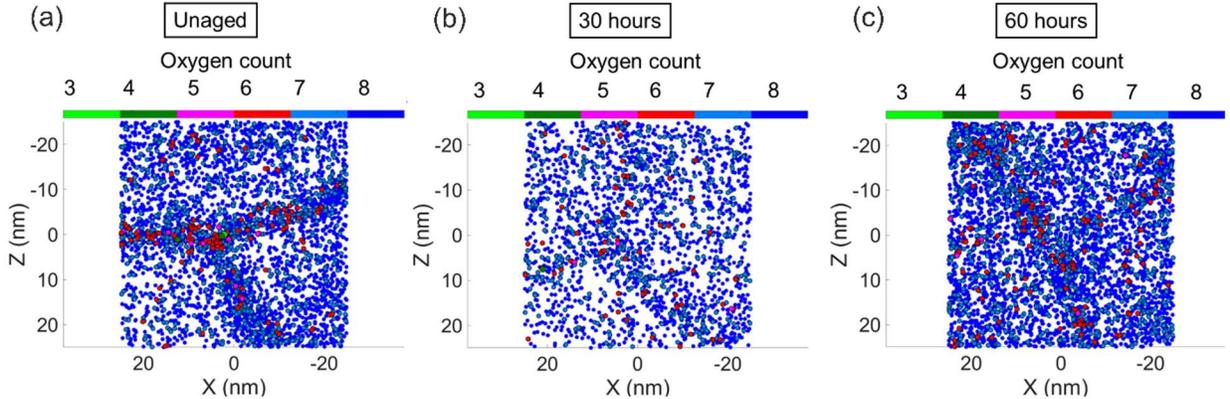

Figure 7. Pseudo-coordination number (PCN) maps showing relative oxygen vacancy distribution in 3 mol% YSZ for (a) unaged, (b) 30 hrs, and (c) 60 hrs aged samples derived from APT. Each panel compares triple junction grain boundaries (TJGBs, 50 × 20 × 50 nm ROIs). Color bars indicate oxygen counts, with lower counts corresponding to higher vacancy concentrations.

## 4. Discussion

Two decades of research on YSZ have reported its exposure to LTD as a limiting factor to its long-term performance, yet the precise mechanisms governing this process remain unexplained. Our innovative machine learning integrated multiscale approach to study controlled hydrothermal aging of YSZ, proposes a clearer view of the LTD mechanism. Because manufacturer specific processing (sintering) conditions are undisclosed yet are known to shift LTD kinetics, the measured monoclinic content naturally differs from the ML predictions. Accordingly, we use ML here as a guide to identify sensitive parameter ranges, not as a batch specific predictor. This aligns with the prior observation that sintering temperature/time affects LTD susceptibility in YSZ. [23, 24] These differences result in a slower rate of t→m phase transformation upon exposure to aqueous environments. Despite this, we successfully captured evidence of the t→m transformation. Following the aging parameters obtained from modeling, samples were aged and analyzed across time. AFM data reveals a reproducible, non-monotonic surface trajectory rather than random scatter or monotonic progress. In stage 1 (0 to 30 hrs), surface roughness increases, and grain size distribution broadens, reflecting t→m plates nucleation at grain boundaries and advancement into adjacent grains. The shear and ~4% volume expansion at these transformation fronts pushes boundaries locally, merging segments and creating the appearance of coarse surface grains at AFM scale. XRD confirms early, near surface monoclinic accumulation, consistent with LTD initiating at free surfaces or grain boundaries. Additionally, AFM roughness maximum in this interval is a signature of relief generated by t→m.

In stage 2, (> 30 hrs), surface roughness reduces while the grain size distribution tightens, evidenced by narrower box plot and shorter whiskers relative to stage 1. This indicates towards a continued transformation coupled with relief redistribution within the already transformed layer. As the transformed region thickens, the surface corrugation reorganizes over a larger region, reducing roughness at the AFM field of view, even as the transformation proceeds subsurface (as seen by XRD's depth gradient). Therefore, apparent refinement of the grain map reflects fragmentation of previously coalesced regions by advancing transformation plates and crack paths rather than a reversal of thermal grain growth. APT provides boundary specific chemistry and defect signatures supporting this model. We observe Al/Si rich intergranular boundaries (AlGBs) originating from unintentional impurities that tend to retain oxygen vacancies, consistent with reports that minor Al additions can slow LTD. In contrast, TJGBs exhibit relative vacancy depletion after prolonged aging as indicated by PCN results. The ionic radius mismatch between $Al^{3+}$ (67.5 pm) and $Zr^{4+}$ (84.0 pm), as well as between $Y^{3+}$ (90.0 pm) and $Zr^{4+}$, results in stronger GB segregation behavior. TJGB are geometrically high energy and highly connected nodes that offer short path lengths for water interactions and defect exchange. Once the surface is partially transformed, the density of junctions per unit area increases, making TJGBs the natural transformation hotspots for continued t→m process. In parallel, segregation controlled differences in boundary structure, like thinner AlGBs vs. Y-enriched GBs, modulate local vacancy stability, explaining why some boundaries resist vacancy loss while TJGBs do not. This mechanism, validated across multiple scales, reveals TJGBs as critical loci of LTD, shifting YSZ from apparent coarsening to 30 hrs, reflecting the transformation plate growth and local boundary motion near the surface. Subsequent partial refinement reflects fragmentation/redistribution as transformation continues, and boundary chemistry evolves. This two-phase model challenges conventional views of LTD as a uniform, surface-to-bulk progression. While prior studies identified various reasons for degradation [7,15,14], our AFM, and APT data pinpoint TJGBs as the dominant weak links, their high vacancy concentrations, and disorder, amplifying hydrothermal effects. The observed vacancy annihilation at TJGBs, contrasted with stable bulk levels, suggests water molecules target these junctions, destabilizing the tetragonal phase.

Together, our multiscale data proposes a transformation coupled junction mediated sequential degradation model as shown in Figure 8:

1. Stage 1 (0 hrs to 30 hrs): t→m nucleates at boundaries, aided by near surface water, and transformation laths grow into grains. The associated volume expansion and shear increase Rq and broadens the size of distribution (apparent coarsening)
2. Stage 2 (30 hrs to 60 hrs): continued transformation plus dopant redistribution led to partial refinement (narrower distribution than at 30 hrs) and lower Rq. As junction density increases, TJGBs, which undergo vacancy depletion, become dominant reaction nodes, sustaining the t→m progress despite vacancy retaining behavior at some AlGBs.

In nanocrystalline ceramics, a transition to boundary-controlled deformation can occur below a critical grain size, producing an inverse Hall Petch like softening where hardness/strength decreases as grains size reduces. This behavior has been directly observed in nanograined YSZ, where both Hall Petch and inverse Hall Petch regimes appear with a transition across tens hundreds of nanometers (e.g., ~55 –250 nm 3YSZ; additional reports of room-temperature softening in nanograined YSZ) and more broadly in other nanocrystalline ceramics [27, 28, 25, 26, 27]. In our commercial 3YSZ, average grains are larger than the canonical nanocrystalline regime, therefore, we note it as a plausible secondary facilitator that is directionally consistent with our kinetics. As the microstructure refines during Stage II, the increased grain boundary and especially TJ area can enhance boundary-mediated transport and local rearrangements that favor the t→m an interpretation that also aligns with known roles of GB/TJGB chemistry and water access in LTD.

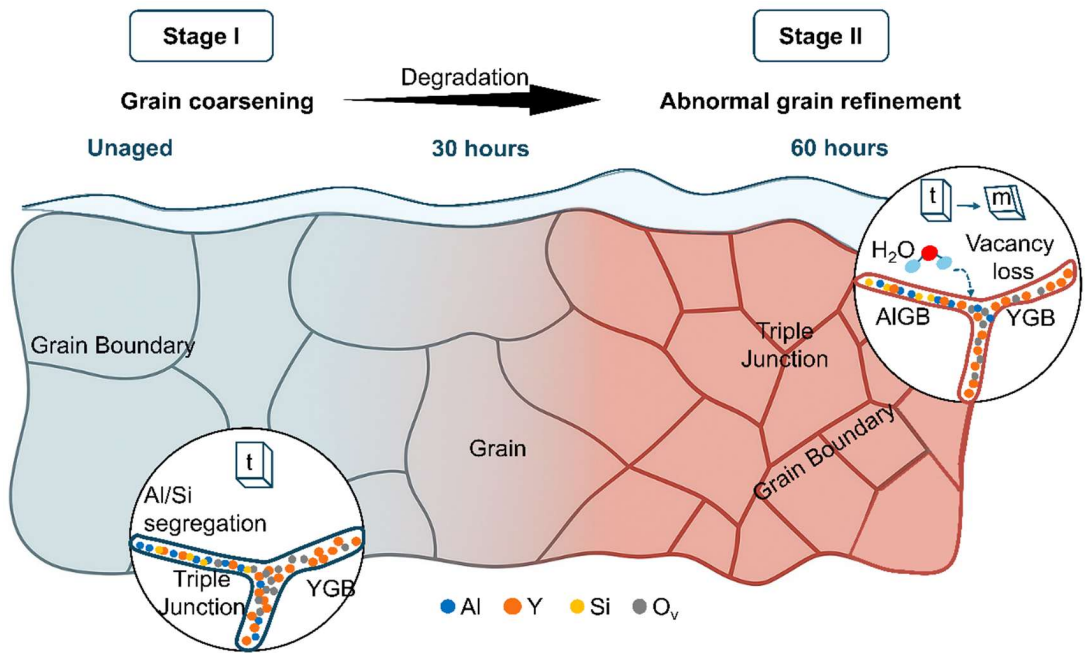

Figure 8: Illustration of the microstructural and chemical evolution of 3 mol% YSZ during aging at low temperature and humid conditions. The Figure shows grain chemistry progression over time (0, 30, and 60 hrs), highlighting initial grain coarsening followed by partial refinement and increased triple junction formation. Insets depict localized phenomena: (left) mostly tetragonal phase with Al and Si segregation at intergranular boundaries (AIGBs) and Y segregation at Y-rich grain boundaries (YGBs), with thinner AIGBs; (right) degradation at triple junction grain boundaries (TJGBs) involving $H_2O$ infiltration, oxygen vacancy loss, and tetragonal-to-monoclinic (t → m) transformation. Notably, YGBs are thicker than AIGBs, and GB thickness decreases with aging, consistent with dopant redistribution and boundary densification.

## 2.1.Implications of degradation mechanisms

This interplay between GB chemistry and LTD revealed in this study has direct implications for YSZ's performance in advanced ceramic applications. Specifically, oxygen vacancy depletion at TJGBs accelerates the t→m transformation, reducing the tetragonal phase is critical for transformation toughening [1]. This phase shift and associated microcracking lower fracture toughness, potentially by 20–30% in biomedical applications, compromising mechanical reliability [28]. In SOFCs, TJGB-driven vacancy loss disrupts oxygen vacancy networks, decreasing ionic conductivity by up to 15% [29], which reduces fuel cell efficiency and lifespan. These effects highlight the importance of this study in identifying TJGBs as prime targets for enhancing YSZ's stability. This junction specific interpretation points to two practical levers for LTD migration: (i) junction management like processing routes that limit TJ density in susceptible size ranges, and (ii) boundary chemistry engineering like promoting AlGB segregation that stabilizes vacancies and hinders water mediated loss. While this study does not include direct measurements of mechanical or electrochemical performance, our multiscale insights establish a mechanistic basis that underscores the urgent need for future studies linking LTD induced microstructural evolution to property degradation in real world applications.

## 5. Conclusion

This multiscale investigation into the degradation mechanism of 3 mol% YSZ under controlled hydrothermal aging reveals the complex interplay between grain boundary chemistry and structure, offering a foundation for enhancing the long-term stability of ceramics. By integrating machine learning-guided aging with mesoscale (AFM, XRD) and atomic-scale (APT) characterization, we identify a two-stage degradation pathway. An initial surface relief building assisted boundary motion presents as apparent coarsening to 30 hrs, followed by partial refinement and relief redistribution from 30 hrs to 60 hrs. XRD confirms a surface initiated increase of monoclinic content with a depth gradient showing LTD progresses inwards over time. APT links this evolution to boundary chemistry and defect transport. TJGBs emerge as dominant sites due to vacancy loss, consistent with water mediated annihilation at highly connected, high-energy regions, aligning with the emergence of monoclinic phases. In contrast, Al/Si segregation at AlGB stabilizes vacancies, sustaining the t→m shift, which suggests locally greater resistance to vacancy depletion compared with YGBs. These findings highlight the critical role of grain boundary chemistry in LTD, with TJGB and AIGB as key sites for degradation. Despite transient surface stabilization, persistent TJGB density poses risks to material reliability. Our results suggest that grain boundary engineering through controlled doping (e.g., reducing Al/Si impurities) or optimizing grain size could mitigate LTD. These insights lay the groundwork for designing more durable YSZ for energy, biomedical, and thermal applications, with future studies encouraged to link these mechanistic findings to performance metrics like toughness and conductivity, translating mechanistic insight into practical design strategies.

## 6. Acknowledgements

This work was supported by the National Science Foundation award number DMR2114595. The authors acknowledge Dr. Prathima Nalam for providing access to atomic force microscopy, Dr. Fei Yao and Dr. Fu Yu for their assistance with hydrothermal aging experiments. The authors also acknowledge US Department of Energy, Office of Science User Facility at Center for Nanophase Materials Sciences (CNMS), Oak Ridge National Laboratory, where APT experiments were conducted by Dr. Jonathon Poplawsky and Dr. James Burns. The authors further acknowledge Dr.

Matthew Brahlek and Dr. An-Hsi Chen for their support with XRD experiments at Oak Ridge National Laboratory.

## 7. Conflict of Interest

The authors declare no conflict of interest.

## Reference

[1] Ramesh, S., Sara Lee K.Y. & Tan, C.Y., "A Review on the Hydrothermal Ageing Behaviour of Y-TZP Ceramics," *Ceramics International* 44, 20620–34, https://doi.org/10.1016/j.ceramint.2018.08.216

[2] Chevalier, J. & Gremillard, L., "Ceramics for Medical Applications: A Picture for the next 20 Years," *Journal of the European Ceramic Society* 29, 1245–55, https://doi.org/10.1016/j.jeurceramsoc.2008.08.025

[3] Anselmi-Tamburini, U., Woolman, J.N. & Munir, Z. A., "Transparent Nanometric Cubic and Tetragonal Zirconia Obtained by High-Pressure Pulsed Electric Current Sintering," *Advanced Functional Materials* 17, 3267–73, https://doi.org/10.1002/adfm.200600959

[4] Ragurajan, D., Golieskardi, M., Satgunam, M. & Min Hwei Ng, A., "Microstructure Analysis and Low Temperature Degradation Resistance of Doped 3Y-TZP Ceramics for Hip Implant Application," *Key Engineering Materials* 706, 42–47, https://doi.org/10.4028/www.scientific.net/KEM.706.42

[5] Daou, E.E., "The Zirconia Ceramic: Strengths and Weaknesses," *The Open Dentistry Journal* 8, 33–42, https://doi.org/10.2174/1874210601408010033

[6] Zhang, F., et al., "3Y-TZP Ceramics with Improved Hydrothermal Degradation Resistance and Fracture Toughness," *Journal of the European Ceramic Society* 34, 2453–63, https://doi.org/10.1016/j.jeurceramsoc.2014.02.026

[7] Chevalier, J., Gremillard, L., Virkar, A.V. & Clarke, D.R., "The Tetragonal-Monoclinic Transformation in Zirconia: Lessons Learned and Future Trends," *Journal of the American Ceramic Society* 92, 1901–20, https://doi.org/10.1111/j.1551-2916.2009.03278.x

[8] Gui, J. & Xie, Z., "Phase Transformation and Slow Crack Growth Study of Y-TZP Dental Ceramic," *Materials Science and Engineering: A* 676, 531–35, https://doi.org/10.1016/j.msea.2016.09.026

[9] Kohorst, P., et al., "Low-Temperature Degradation of Different Zirconia Ceramics for Dental Applications," *Acta Biomaterialia* 8, 1213–20, https://doi.org/10.1016/j.actbio.2011.11.016

[10] Matsui, K., Nakamura, K., Kumamoto, K., Yoshida, H. & Ikuhara, Y., "Low-Temperature Degradation in Yttria-Stabilized Tetragonal Zirconia Polycrystal Doped with Small Amounts of Alumina: Effect of Grain-Boundary Energy," *Journal of the European Ceramic Society* 36, no. 1 (January 2016): 155–62, https://doi.org/10.1016/j.jeurceramsoc.2015.09.016

[11] Reddy, A.S.T., Balaji, N. & Rajam, K.S., "Phase Transformation and Wear Studies of Plasma Sprayed Yttria Stabilized Zirconia Coatings Containing Various Mol% of Yttria," *Materials Characterization* 62, 697–705, https://doi.org/10.1016/j.matchar.2011.04.018

[12] Kawai, Y., Uo, M., Wang, Y., Kono, S., Ohnuki & Watari, F., "Phase Transformation of Zirconia Ceramics by Hydrothermal Degradation," *Dental Materials Journal* 30, 286–92, https://doi.org/10.4012/dmj.2010-175

[13] Kern, F., "Evidence of Phase Transitions and Their Role in the Transient Behavior of Mechanical Properties and Low Temperature Degradation of 3Y-TZP Made from Stabilizer-Coated Powder," *Ceramics* 2, 271–85, https://doi.org/10.3390/ceramics2020022

[14] Hallmann, L., et. al. "The Influence of Grain Size on Low-temperature Degradation of Dental Zirconia," *Journal of Biomedical Materials Research Part B: Applied Biomaterials* 100B, 447–56, https://doi.org/10.1002/jbm.b.31969

[15] Zhang, F., et al., "Effect of Cation Dopant Radius on the Hydrothermal Stability of Tetragonal Zirconia: Grain Boundary Segregation and Oxygen Vacancy Annihilation," *Acta Materialia* 106, 48–58, https://doi.org/10.1016/j.actamat.2015.12.051

[16] Li, P., Chen, I.W. & Penner-Hahn, J.E., "Effect of Dopants on Zirconia Stabilization—An X-ray Absorption Study: II, Tetravalent Dopants," *Journal of the American Ceramic Society* 77, 1281–88, https://doi.org/10.1111/j.1151-2916.1994.tb05403.x.

[17] Zhu, W., Nakashima, S., Marin, E., Gu, H. & Pezzotti, G., "Microscopic Mapping of Dopant Content and Its Link to the Structural and Thermal Stability of Yttria-Stabilized Zirconia Polycrystals," *Journal of Materials Science* 55, 524–34, https://doi.org/10.1007/s10853-019-04080-9

[18] Zhou, Q., Liu, Y., Ju, W., Zhang, Q. & Li., J, "Exploring the Effect of Dopant (Si, P, S, Ge, Se, and Sb) in Arsenene: A DFT Study," *Physics Letters A* 384, 126146, https://doi.org/10.1016/j.physleta.2019.126146.

[19] Feng, B., Lugg, N.R., Kumamoto, A., Ikuhara, Y. & Shibata, N., "Direct Observation of Oxygen Vacancy Distribution across Yttria-Stabilized Zirconia Grain Boundaries," *ACS Nano* 11, 11376–82, https://doi.org/10.1021/acsnano.7b05943

[20] An, J., et al., "Atomic Scale Verification of Oxide-Ion Vacancy Distribution near a Single Grain Boundary in YSZ," *Scientific Reports* 3, 2680, https://doi.org/10.1038/srep02680

[21] Jue, J.F., Chen, J. & Virkar, A.V., "Low-Temperature Aging of $t\,'$-Zirconia: The Role of Microstructure on Phase Stability," *Journal of the American Ceramic Society* 74, 1811–20, https://doi.org/10.1111/j.1151-2916.1991.tb07793.x

[22] Licata, O.G., Zhu, M., Hwang, J. & Mazumder, B., et al., "Nanoscale Chemistry and Ion Segregation in Zirconia-Based Ceramic at Grain Boundaries by Atom Probe Tomography," *Scripta Materialia* 213, 114603, https://doi.org/10.1016/j.scriptamat.2022.114603

[23] Jian Wang et al., "Effect of Sintering Temperature on Phase Transformation Behavior and Hardness of High-Pressure High-Temperature Sintered 10 Mol% Mg-PSZ," *Ceramics International* 47, no. 11 (2021): 15180–85, https://doi.org/10.1016/j.ceramint.2021.02.078.

[24] Qiannan Li et al., "Controlled Sintering and Phase Transformation of Yttria-Doped Tetragonal Zirconia Polycrystal Material," *Ceramics International* 47, no. 19 (2021): 27188–94, https://doi.org/10.1016/j.ceramint.2021.06.139.

[25] Lin Feng et al., "Size-Induced Room Temperature Softening of Nanocrystalline Yttria Stabilized Zirconia," *Journal of the European Ceramic Society* 40, no. 5 (2020): 2050–55, https://doi.org/10.1016/j.jeurceramsoc.2020.01.046.

[26] Heonjune Ryou et al., "Below the Hall–Petch Limit in Nanocrystalline Ceramics," *ACS Nano* 12, no. 4 (2018): 3083–94, https://doi.org/10.1021/acsnano.7b07380.

[27] Saeed Zare Chavoshi et al., "Transition between Hall-Petch and Inverse Hall-Petch Behavior in Nanocrystalline Silicon Carbide," *Physical Review Materials* 5, no. 7 (2021): 073606, https://doi.org/10.1103/PhysRevMaterials.5.073606.

[28] Guo, X., "Hydrothermal Degradation Mechanism of Tetragonal Zirconia," Journal of Material Science 36, 3737 – 3744, https://doi.org/10.1023/A:1017925800904

[29] Guo, X. & Maier, J., "Grain Boundary Blocking Effect in Zirconia: A Schottky Barrier Analysis," *Journal of The Electrochemical Society* 148, E121, https://doi.org/10.1149/1.1348267

Supplementary information

Revealing degradation mechanisms in YSZ ceramics through machine learning-guided aging and multiscale characterization

Prachi Garg1, and Baishakhi Mazumder1*
1 Department of Materials Design and Innovation, State University of New York at Buffalo, USA
*baishakh@buffalo.edu (corresponding author)

Literature data for ML model training to optimize aging parameter

Literature-based aging parameters were used to train predictive regression models. Around 200 values were extracted from the literature. To ensure the model's reliability, the repetitive values were removed. As a result, the model was trained on around 100 values. Each value is provided with a reference in Table S1.

Table S1. List of aging parameters extracted from literature.

| Monoclinic phase (%) | Temp (c) | Time (hrs) | Pressure (bar) | Reference |
|---|---|---|---|---|
| 12 | 134 | 0 | 3 | |
| 45 | 134 | 4 | 3 | |
| 55 | 134 | 8 | 3 | |
| 65 | 134 | 16 | 3 | [1] |
| 70 | 134 | 32 | 3 | |
| 73 | 134 | 64 | 3 | |
| 73 | 134 | 128 | 3 | |
| 11 | 200 | 16 | 6 | |
| 0 | 134 | 40 | 2 | |
| 10 | 134 | 40 | 2 | |
| 80 | 140 | 48 | 3.6 | |
| 2.8 | 180 | 200 | 10 | |
| 88 | 180 | 60 | 10 | |
| 35 | 140 | 20 | 2 | |
| 78 | 134 | 48 | 2 | |
| 73.6 | 140 | 168 | 4 | |
| 0 | 134 | 10 | 2 | [2] |
| 0 | 134 | 40 | 2 | |
| 3 | 140 | 6 | 3.6 | |
| 45 | 134 | 48 | 2 | |
| 10 | 134 | 40 | 2 | |
| 70 | 140 | 48 | 3.6 | |
| 80 | 100 | 72 | 1 | |
| 0 | 134 | 25 | 3 | |
| 0 | 245 | 168 | 4 | |
| 10 | 134 | 40 | 2 | |
| 73.6 | 140 | 4 | 168 | |
| 0 | 134 | 1 | 3 | [3] |
| 23 | 134 | 4 | 3 | |
| 29 | 134 | 8 | 3 | |
| 46 | 134 | 16 | 3 | |
| 69 | 134 | 32 | 3 | |
| 75 | 134 | 64 | 3 | |
| 76 | 134 | 128 | 3 | |
| 5.58 | 134 | 0 | 2.07 | [4] |
| 34.7 | 134 | 49 | 2.07 | |
| 0 | 132 | 0 | 2 | [5] |
| 0.39 | 132 | 5 | 2 | |
| 134 | 0 | 2 | 0 | (6) |
| 134 | 1 | 2 | 0.5 | |
| 134 | 3 | 2 | 6.4 | |
| 134 | 5 | 2 | 13.7 | |
| 134 | 10 | 2 | 38.5 | |
| 2.13 | 121 | 1 | 1 | [7] |
| 6.3 | 121 | 3 | 1 | |
| 9.27 | 121 | 5 | 1 | |
| 2.92 | 134 | 1 | 2 | |
| 12.6 | 134 | 3 | 2 | |
| 18.1 | 134 | 5 | 2 | |
| 0 | 134 | 0 | 2 | [8] |
| 60.47 | 134 | 20 | 2 | |
| 1.37 | 127 | 0 | 1.5 | (9) |
| 23.4 | 127 | 12 | 1.5 | |
| 0 | 103 | 0 | 2.07 | |
| 23.2 | 103 | 6 | 2.07 | |
| 42.9 | 103 | 20 | 2.07 | |
| 72 | 103 | 40 | 2.07 | |
| 80.6 | 103 | 60 | 2.07 | |
| 81.8 | 103 | 138 | 2.07 | |
| 0 | 103 | 0 | 2.07 | |
| 36.1 | 103 | 6 | 2.07 | |
| 53.9 | 103 | 20 | 2.07 | |
| 66.1 | 103 | 40 | 2.07 | |
| 71 | 103 | 60 | 2.07 | |
| 73.3 | 103 | 138 | 2.07 | |
| 0 | 103 | 0 | 2.07 | |

| | | | |
|---|---|---|---|
| 30.5 | 103 | 6 | 2.07 |
| 49.9 | 103 | 20 | 2.07 |
| 57.3 | 103 | 40 | 2.07 |
| 59.9 | 103 | 60 | 2.07 |
| 63.2 | 103 | 138 | 2.07 |
| 0.3 | 121 | 0 | 2 |
| 2 | 134 | 0 | 3 |
| 7 | 134 | 8 | 3 |
| 0.4 | 134 | 0 | 1 |
| 31 | 134 | 24 | 1 |
| 64 | 134 | 96 | 1 |
| 68 | 134 | 168 | 1 |
| 0 | 134 | 0 | 2 |
| 30 | 134 | 12 | 2 |
| 13.35 | 134 | 0 | 2 |
| 25.4 | 134 | 15 | 2 |
| 1.02 | 134 | 0 | 2 |
| 14.5 | 134 | 6 | 2 |
| 13.5 | 134 | 4 | 2 |
| 41 | 134 | 20 | 2 |
| 54.7 | 134 | 30 | 2 |
| 64.5 | 134 | 40 | 2 |
| 0 | 134 | 0 | 2 |
| 0 | 122 | 0 | 2 |
| 55 | 122 | 20 | 2 |
| 0 | 121 | 0 | 2 |
| 5 | 121 | 5 | 2 |
| 11 | 121 | 10 | 2 |

Reference:

1.      Kohorst, P., et al. "Low-temperature degradation of different zirconia ceramics for dental applications." Acta Biomaterialia 8, 1213–20,  http://dx.doi.org/10.1016/j.actbio.2011.11.016

2.      Ramesh, S., Sara, Lee, K.Y., Tan, C.Y., "A review on the hydrothermal ageing behaviour of Y-TZP ceramics." Ceramics International 44, 20620–34, http://dx.doi.org/10.1016/j.ceramint.2018.08.216

3.      Hübsch, C., Dellinger, P., et al. "Protection of yttria-stabilized zirconia for dental applications by oxidic PVD coating." Acta Biomaterialia 11, 488–93, http://dx.doi.org/10.1016/j.actbio.2014.09.042

4.      Roy, M.E., Whiteside, L.A., Katerberg, B.J., Steiger, J.A., "Phase transformation, roughness, and microhardness of artificially aged yttria- and magnesia-stabilized zirconia femoral heads." Journal of Biomedical Material Research: A 83, 1096-1102, http://doi.org/10.1002/jbm.a.31438

5. Lucas, T.J., Lawson, N.C., Janowski, G.M., Burgess, J.O., "Phase transformation of dental zirconia following artificial aging." Journal of Biomedical Material Research 103, 1519–23, http://doi.org/10.1002/jbm.b.33334

6. Sevilla, P., Sandino, C., Arciniegas, M., Martínez-Gomis, J., Peraire, M., Gil, F.J, "Evaluating mechanical properties and degradation of YTZP dental implants." Materials Science and Engineering: C 30, 14–9, http://doi.org/10.1016/j.msec.2009.08.002

7. Lucas, T.J., Lawson, N.C., Janowski, G.M., Burgess, J.O., "Effect of grain size on the monoclinic transformation, hardness, roughness, and modulus of aged partially stabilized zirconia." Dental Materials 31, 1487–92, http://dx.doi.org/10.1016/j.dental.2015.09.014

8. Silvestri, T., Pereira, G.K.R., Guilardi, L.F., Rippe, M.P., Valandro, L.F., "Effect of Grinding and Multi-Stimuli Aging on the Fatigue Strength of a Y-TZP Ceramic." Brazilian Dental Journal 29, 60–7, http://dx.doi.org/10.1590/0103-6440201801735

9. Pereira, G.K.R., et al. "Low-temperature degradation of Y-TZP ceramics: A systematic review and meta-analysis." Journal of the Mechanical Behavior of Biomedical Materials 55, 151–63, http://dx.doi.org/10.1016/j.jmbbm.2015.10.017